\documentclass[aps,prd,11pt,notitlepage,nofootinbib,superscriptaddress,showkeys,showpacs]{revtex4-1}
\usepackage{amsmath,amssymb,amsthm,latexsym}
\usepackage{stmaryrd,wasysym,upgreek,mathrsfs,dsfont}
\usepackage[english]{babel}
\usepackage{graphicx,color}
\usepackage{xspace}
\usepackage{graphicx}
\usepackage{pifont,dsfont}
\usepackage{marvosym}
\usepackage{slashed}
\usepackage{subfigure}

\newcommand{\be}{\begin{equation}}
\newcommand{\ee}{\end{equation}}
\newcommand{\bqa}{\begin{eqnarray}}
\newcommand{\eqa}{\end{eqnarray}}
\newcommand{\bea}{\begin{eqnarray}}
\newcommand{\eea}{\end{eqnarray}}

\newcommand{\Tr}{{\rm Tr}}

\newtheorem{lemma}{Lemma}

\newtheorem{definition}{Definition}

\newtheorem{remark}{Remark}

\newtheorem{proposition}{Proposition}

\newcommand{\cF}{{\cal F}}

\newcommand{\cG}{{\cal G}}
\newcommand{\cB}{{\cal B}}
\newcommand{\cJ}{{\cal J}}
\newcommand{\cH}{{\cal H}}
\newcommand{\cM}{{\cal M}}
\newcommand{\cN}{{\cal N}}

\DeclareMathOperator{\tr}{tr}
\DeclareMathOperator{\vol}{vol}
\DeclareMathOperator{\Vol}{Vol}

\begin{document}

\title{\Large \bf Critical behavior of colored tensor models in the large $N$ limit}

\author{{\bf Valentin Bonzom}}\email{vbonzom@perimeterinstitute.ca}
\author{{\bf Razvan Gurau}}\email{rgurau@perimeterinstitute.ca}
\affiliation{Perimeter Institute for Theoretical Physics, 31 Caroline St. N, ON N2L 2Y5, Waterloo, Canada}

\author{{\bf Aldo Riello}}  \email{aldo.riello@ens.fr}
\author{{\bf Vincent Rivasseau}}  \email{vincent.rivasseau@gmail.com}
\affiliation{Laboratoire de Physique Th\'eorique, CNRS UMR 8627, Universit\'e Paris XI,  F-91405 Orsay Cedex, France}

\date{\small\today}

\begin{abstract}
Colored tensor models have been recently shown to admit a large $N$ expansion, whose leading order encodes
a sum over a class of colored triangulations of the $D$-sphere. The present paper investigates in details
this leading order. We show that the relevant triangulations proliferate
like a species of colored trees. The leading order is therefore summable and exhibits a critical behavior,
independent of the dimension. A continuum limit is reached by tuning the coupling constant to its critical
value while inserting an infinite number of pairs of $D$-simplices glued together in a specific way.
We argue that the dominant triangulations are branched polymers.
\end{abstract}

\medskip

\noindent  Pacs numbers: 02.10.Ox, 04.60.Gw, 05.40-a

\maketitle

\section{Introduction}

Tensor models  \cite{mmgravity,ambj3dqg,sasa1} and group field theories
\cite{Boul,Ooguri:1992eb,laurentgft,quantugeom2} are the natural generalization of matrix models \cite{mm,Di Francesco:1993nw}
implementing in a consistent way the sum over random triangulations in dimensions higher than two.
They are notoriously hard to control analytically and one usually resorts to
numerical simulations \cite{Ambjorn:1991wq,Ambjorn:2000dja,Ambjorn:2005qt}.
Progress has recently been made in the analytic control of tensor models with the advent of the
$1/N$ expansion \cite{Gur3,GurRiv,Gur4} of {\em colored} \cite{color,lost,PolyColor} tensor models. This expansion
synthetizes several alternative evaluations of graph amplitudes in tensor models
\cite{FreiGurOriti,sefu1,sefu2,sefu3,BS1,BS2,Geloun:2011cy,BS3,OC}
and provides a straightforward generalization of the familiar genus expansion of matrix models
\cite{'tHooft:1973jz,Brezin:1977sv} in arbitrary dimension.
The coloring of the fields allows one to address previously inaccessible questions in tensor models like the
implementation of the diffeomorphism symmetry
\cite{Baratin:2011tg,OC} in the Boulatov model or the identification of embedded matrix models \cite{Ryan:2011qm}.
The symmetries of tensor models have recently been studied using n-ary algebras
\cite{Sasakura:2011ma,Sasakura:2011nj}.

This paper is the first in a long series of studies of the implications of the $1/N$ expansion
in colored tensor models. We present here a complete analysis of the leading order in the large $N$ limit
in arbitrary dimensions, indexed by graphs of spherical topology \cite{GurRiv}. To perform the study
of this leading order one needs to address the following two questions
\begin{itemize}
 \item What is the combinatorics of the Feynman graphs contributing to the leading order, i.e.
  the higher dimensional extension of the notion of planar graphs? Unlike in matrix models, where
  planarity and spherical topology are trivially related, this question is non trivial in tensor models.
  In particular {\it not} all triangulations of the sphere contribute to the leading order.
 \item Is the series of the leading order summable with a non zero radius of convergence? If this is the case,
  then, in the large $N$ limit, the model exhibits a critical behavior whose critical exponents one needs to
  compute.
\end{itemize}

   The two points are intimately intertwined. Indeed, were all the triangulations of the sphere
to contribute to the leading order, one would have difficulties to control its sum: the dependence of the number
of triangulations of the sphere on the number of simplices is, to our knowledge, still a open issue in combinatorial topology.

   We will deal in this paper with the independent identically distributed (i.i.d.) colored tensor model.
As the analysis we perform is strictly combinatorial, the results we obtain have some universality. Not only the tensor
indices need not be simple integers (they can for instance index the
Fourier modes of an arbitrary compact Lie group, or even of a finite group of large order \cite{bianca-finitegp}), but we also expect similar conclusions for the colored Boulatov-Ooguri model,
as the latter supports a $1/N$ expansion dominated by the same family of graphs \cite{Gur4}.

The most natural interpretation of the i.i.d. model is as a model of dynamical triangulations. A graph $\mathcal{G}$
in the Feynman expansion of the free energy corresponds to a triangulation $\Delta$ of a pseudo-manifold and
has amplitude
\be
A(\cG) = e^{\kappa_{D-2}N_{D-2}\, -\, \kappa_D N_D},
\ee
where $N_D(\Delta), N_{D-2}(\Delta)$ are the numbers of $D$-simplices and $(D-2)$-simplices of the triangulation.
 The usual parameters of dynamical triangulations are related to the large parameter $N$ and to the coupling $g$
of the tensor model by: $\kappa_{D-2} = \ln N$ and $\kappa_D = \frac12\Bigl( \frac12\,D(D-1)\, \ln N - \ln g\Bigr)$.

The main results of the present paper are
\begin{itemize}
 \item the dominant triangulations can be obtained from an elementary building block, formed by a pair of $D$-simplices which are glued together along $D$ faces (we will loosely refer to these block as melons, and give a precise definition later). Since
each simplex has one free face, one can glue melons together or insert melons into melons.
 \item the melonic triangulations are in one to one correspondence to a family of decorated trees. Since counting
  trees is a well understood problem, we will prove that the free energy of melonic triangulations has finite radius
  of convergence $g=g_c$.
 \item when the coupling approaches the critical value $g_c$, the free energy exhibits a critical behavior
 $F \propto (g_c - g)^{2-\gamma}$ with $\gamma = \frac12$. This leads to a continuous phase
 of large $D$-spheres triangulated by an infinite number of melons. We argue that this phase corresponds to the branched polymer
 phase of dynamical triangulations.
\end{itemize}

This paper is organized as follows. We recall in section \ref{sec:color} the colored matrix and tensor models.
 In section \ref{sec:melons}, we identify the leading sector (with its melons), which we resum in section
 \ref{sec:trees}. Finally, we discuss in section \ref{sec:critical} the critical behavior and its interpretation
in terms of dynamical triangulations for the i.i.d. model.

Throughout this paper, unless otherwise specified, ``graph'' will always mean connected stranded colored graph.

\section{Colored Tensor Models} \label{sec:color}

We start by briefly recalling the classical results concerning the continuum limit in matrix models.
The planar contribution to the free energy of a matrix model with quartic interaction
\cite{Brezin:1977sv}
\be
F_{\rm planar} = - \lim_{N\to \infty} \frac{1}{N^2}\, \ln \Big{(}
\int d^{N^2} M\  e^{-  \bigl[ \Tr \frac{1}{2}  M^2  + \Tr \frac{g}{N} M^4  \bigr]}
 \Big{)} \; ,
\ee
computes to
\bea  F_{\rm planar} (g) = \frac{1}{24}  (a^2 - 1 ) (9- a^2 )-  \frac{1}{2} \ln a^2 , \qquad
a^2 = \frac{1}{24 g}  \bigl( \sqrt{1 + 48 g} -1 \bigr) \; .
\eea
Expanding $F_{\rm planar}$ in $g$, the general term has an asymptotic behavior
$(-48g)^n  n^{-7/2}$, hence the planar sector has radius of convergence
$g_c = -1/48$. The free energy exhibits a critical behavior
$F(g) \sim (g_c-g)^{2-\gamma}$ with\footnote{That exponent is known as the string susceptibility exponent, or
as the entropy exponent: it characterizes asymptotically the number of planar graphs with a fixed number
of vertices.} $\gamma = -1/2$,
corresponding to pure gravity in $D=2$ \cite{Di Francesco:1993nw}. A model with a generic interaction and
 all coupling constants positive falls in the same universality class (i.e. same critical exponents, though
the critical values of the couplings may differ).
The colored matrix model \cite{difrancesco-rect} is
defined by the partition function
\begin{gather}
\nonumber Z = \int dM_1 dM_2 dM_3\quad e^{-N \tr( V(M_1,M_2,M_3))},\\
V(M_1,M_2,M_3) = \frac1\tau\left(M_1\,M_1^\dagger + M_2\,M_2^\dagger + M_3\,M_3^\dagger \right)
- M_1M_2M_3 - M_3^\dagger M_2^\dagger M_1^\dagger \; .
\end{gather}
Its perturbative expansion is indexed by line colored graphs. In two dimensions the
coloring of the lines can be translated
into a coloring of the faces of the graph, hence a coloring of the vertices in the dual
triangulation, \cite{difrancesco-rect}. Observing that the model is Gaussian in two colors,
one can rewrite it as a one-matrix model with non-polynomial potential
 $V'(M) = \tr\left( \frac{M M^\dagger}{\tau^3} + \log (I - MM^\dagger)\right)$, which, as all its
 couplings are positive, falls also in the universality class of pure
two-dimensional quantum gravity \cite{difrancesco-coloringRT, difrancesco-countingRT}.
Note that \cite{difrancesco-countingRT} offers a combinatorial proof of this statement, by
mapping colored planar graphs onto a family of decorated trees, quite close in spirit to
our derivation.

\subsection{The independent identically distributed colored tensor model}

We denote $\vec n_i$, for $i=0,\dotsc,D$, the $D$-uple of integers
 $\vec n_i = (n_{ii-1},\dotsc, n_{i0},\; n_{iD}, \dotsc, n_{ii+1}) $, with
$n_{ik}=1,\dotsc, N$. This $N$ is the size of the tensors and the large $N$ limit defined in
\cite{Gur3, GurRiv, Gur4} represents the limit of infinite size tensors.
We set $n_{ij} = n_{ji}$. Let $\bar \psi^i_{\vec n_i},\; \psi^i_{\vec n_i}$, with $i=0,\dotsc, D$, be $D+1$ couples of complex
conjugated tensors with $D$ indices. The independent identically distributed (i.i.d.) colored tensor
model in dimension $D$ \cite{color,lost,Gur4} is defined by the partition function
\begin{gather}
\nonumber e^{ N^D F_N(\lambda,\bar\lambda)} = Z_N(\lambda, \bar{\lambda}) = \int \, d\bar \psi \, d \psi
\ e^{- S (\psi,\bar\psi)} \; , \\
S (\psi,\bar\psi) = \sum_{i=0}^{D} \sum_{\vec n} \bar \psi^i_{\vec n_i} \psi^i_{\vec n_i}  +
\frac{\lambda}{ N^{D(D-1)/4} } \sum_{\vec n} \prod_{i=0}^D \psi^i_{ \vec n_i } +
\frac{\bar \lambda}{ N^{D(D-1)/4} } \sum_{\vec n}
\prod_{i=0}^D \bar \psi^i_{ \vec n_i } \; . \label{eq:iid}
\end{gather}
$\sum_{\vec n}$ denotes the sum over all indices $\vec n_i$ from $1$ to $N$.
Note that rescaling $\psi^i_{\vec n_i} = N^{-D/4} T^i_{\vec n_i}$ leads to
\bea
 S (\bar T ,T  ) = N^{D/2} \Big{(}
\sum_{i=0}^{D} \sum_{\vec n} \bar T^i_{\vec n_i} T^i_{\vec n_i}  +
\lambda \sum_{\vec n} \prod_{i=0}^D T^i_{ \vec n_i } +
\bar \lambda \sum_{\vec n} \prod_{i=0}^D \bar T^i_{ \vec n_i } \Big{)} \;.
\eea
The partition function of equation \eqref{eq:iid} is evaluated by colored stranded Feynman graphs
\cite{color,lost}. The colors $i$ of the fields $\psi^i, \bar{\psi}^i$ induce important
restrictions on the combinatorics of stranded graphs. Note that we have two types of vertices,
say one of positive (involving $\psi$) and one of negative (involving $\bar \psi$) orientation.
The lines always join a $\psi^i$ to a $\bar{\psi}^i$ and possess a color index.
The tensor indices $n_{jk}$ are preserved along the strands. The amplitude of a graph with $2p$
vertices and $\cF$ faces is \cite{Gur4}
\bea\label{eq:ampli}
A(\cG)= (\lambda\bar\lambda)^p N^{-p \frac{D(D-1)}{2}+ \cF } \; .
\eea

Any Feynman graphs $\cG$ of this model is a simplicial pseudo manifold \cite{lost}. Therefore the colored
tensor models provide a statistical theory of random triangulations in dimensions $D$, generalizing random matrix models.
The vertices, edges and faces (closed strands) of the graph represent the $D$, $(D-1)$ and $(D-2)$-simplices.
The lower dimensional simplices of the pseudo-manifold are identified by the {\bf $n$-bubbles} of the
graph (the maximally connected subgraphs made of lines with $n$ fixed colors).
Bubbles are local objects (in the sense that each of them corresponds to a single simplex of the triangulation)
and encode the cellular structure of the pseudo manifold.  The $0$-bubbles,
$1$-bubbles and $2$-bubbles of a graph are its vertices, lines and faces. Of particular importance in the sequel
are the $D$-bubbles of the graph (that is the maximally connected subgraphs containing
all but one of the colors).  They are associated to the $0$ simplices (vertices) of the pseudo-manifold.
We label $\cB^{\widehat{i}}_{(\rho)}$
the $D$-bubbles with colors $\{0,\dots, D\} \setminus \{i \}$ (and $\rho$ labels the various bubbles
 with identical colors). We denote $\cB^{[D]}$ the total number of $D$ bubbles, which respects
\cite{Gur4}
\be
 p + D - \cB^{[D]} \ge 0 \;,
\ee
where $p$ is half the number of vertices.

A second class of graphs which are crucial for the $1/N$ expansion of the colored tensor model are the
{\bf jackets} \cite{sefu3,Gur3,GurRiv,Gur4}.
\begin{definition}
Let $\tau$ be a cycle on $\{0,\dotsc,D\}$. A colored {\bf jacket} $\cJ$ of $\cG$ is the ribbon graph made by faces
with colors $(\tau^q(0),\tau^{q+1}(0))$, for $q=0,\dotsc,D$, modulo the orientation of the cycle.
\end{definition}
A jacket $\cJ$ of $\cG$ contains all the vertices and all the lines of $\cG$ (hence $\cJ$ and $\cG$ have the same
connectivity), but only a subset of faces. As such, any jacket $\cJ$ carries some key topological information
about $\cG$: for instance the fundamental group of $\cG$ is a subgroup of the fundamental group of any
of its jackets \cite{BS3}. In $D=3$ the jackets of the graph correspond to Heegaard splitting surfaces \cite{Ryan:2011qm}.

Jackets are ribbon graphs (the unique jacket for $D=2$ is the graph itself) hence are completely
classified by their genus $g_\cJ$. For a colored graph $\cG$ we define its {\it degree}
\begin{definition}
The {\bf degree} $\omega(\cG)$ of a graph is the sum of genera of its jackets, $\omega(\cG)=\sum_{\cJ} g_{\cJ}$.
\end{definition}

In $D=2$, the degree is the genus of the ribbon graph. In $D=3$, one can show that the degree is related to the Euler characteristic of the graph
$\omega(\cG) = p-\cB^{[3]} +3 -\chi(\cG)$.

The $D$-bubbles of a graph are also colored graphs,
hence they possess a degree. The two important properties of the degree of a graph are \cite{GurRiv, Gur4}
\bea\label{eq:propdegree}
  &&  \omega(\cG) = \frac{(D-1)!}{2} \Big{(} p+D-\cB^{[D]} \Big{)} + \sum_{i;\rho} \omega(\cB^{\widehat{i}}_{(\rho)}) \; .
   \crcr
  && \frac{2}{(D-1)!} \omega(\cG) = \frac{D (D-1)}{2} p + D - \cF \; .
\eea
The $1/N$ expansion of the colored tensor model is encoded in the remark that $\omega(\cG)$ has exactly
the combination of $p$ and $\cF$ appearing in the amplitude of a graph \eqref{eq:ampli}, thus
\be
A(\cG) = (\lambda\bar\lambda)^p\ N^{D - \frac{2}{(D-1)!} \omega(\cG)  }\;.
\ee
The free energy $F_N(\lambda, \bar{\lambda})$ of the model admits then an expansion in the degree
\be
F_N(\lambda,\bar\lambda) = C^{[0]}( \lambda,\bar\lambda ) + O(N^{-1} ) \;,
\ee
where $C^{[0]}( \lambda,\bar\lambda ) $ is the sum over all graphs of degree $0$.
This paper is dedicated to the detailed analysis of $C^{[0]}( \lambda,\bar\lambda )$.
The degree plays in dimensions $D\geq 3$ the role played by the genus in matrix models, and in
particular degree $0$ graphs are spheres \cite{Gur4}.

\begin{proposition} \label{planar jacket}
If the degree vanishes (i.e. all jackets of $\cG$ are planar) then $\cG$ is dual to a $D$-sphere.
\end{proposition}

We end this section with the derivation of the simplest Schwinger-Dyson equation of the i.i.d. colored tensor model.
Due to the conservation of the indices along the strands,
the two point function is necessarily connected and has the index structure
\bea\label{eq:2point}
\langle \bar{\psi}^i_{\vec{n}_i}\,\psi^i_{\vec{p}_i}\rangle_{\rm c} =
  \delta_{\vec{n}_i,\vec{p}_i} \,G_N(\lambda,\bar{\lambda}) \; ,
\eea
where $\delta_{\vec{n}_i,\vec{p}_i} $ denotes $ \prod_{k\neq i} \delta_{n_{ik} p_{ik}}$.
\begin{proposition} \label{relation GF}
The full connected 2-point function is
\be
G_N(\lambda,\bar{\lambda}) = 1 +\lambda\,\partial_\lambda F_N(\lambda,\bar{\lambda}) \; ,
\ee
where $F_N$ is the free energy of the model.
\end{proposition}

\noindent{\bf Proof:} We start from the trivial identity
\bea
\frac{1}{Z_N(\lambda, \bar\lambda)} \int d\bar \psi \, d\psi\  \frac{\delta}{\delta \psi^i_{\vec n_i }} \Big{(}
  \psi^i_{\vec p_i} \, e^{-S(\psi,\bar\psi)}
\Big{)} =0 \; ,
\eea
which computes
\begin{multline}
\delta_{\vec{n}_i, \vec{p}_i} - \frac{1}{Z_N(\lambda, \bar\lambda)} \int d\bar \psi \, d\psi \;
  \psi^i_{\vec p_i} \,\bar \psi^i_{\vec n_i}\ e^{-S(\psi,\bar\psi)}\\
- \frac{\lambda}{N^{D(D-1)/4 }} \frac{1}{Z_N(\lambda, \bar\lambda)} \int d\bar \psi \, d\psi \;
\psi^i_{\vec p_i} \sum_{n \neq n_{ji} } \prod_{j\neq i} \psi^j_{\vec n_j}\ e^{-S(\psi,\bar\psi)} = 0 \; .
\end{multline}
Setting $\vec p_i =\vec n_i$ and summing over $\vec n_i$ we get
\bea
&& N^D - N^D  \,G_N(\lambda,\bar{\lambda}) - \frac{\lambda}{N^{D(D-1)/4 }}
\frac{1}{Z_N(\lambda, \bar\lambda)} \int d\bar \psi \, d\psi \;
 \Big{(}\sum_{n } \prod_{i} \psi^i_{\vec n_i}\Big{)} e^{-S}  \crcr
&&= N^D - N^D  \,G_N(\lambda,\bar{\lambda}) + \lambda \frac{1}{ Z_N(\lambda, \bar\lambda) }
\partial_{\lambda} Z_N(\lambda, \bar\lambda)
=0 \;,
\eea
which, recalling that $ F_{N}(\lambda,\bar\lambda) = N^{-D} \ln  Z_N(\lambda, \bar\lambda) $ proves the lemma.

\qed

\section{The dominant order: the world of melons} \label{sec:melons}

\begin{figure}[htb]
\subfigure[The ``super melon'' graph.]{
 \includegraphics[width=2cm]{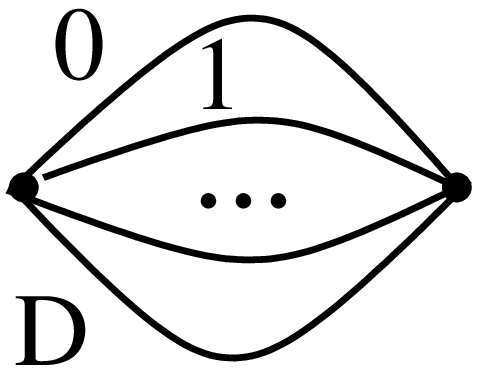}
  \label{fig:supermelon}
     }
 \hskip3cm
 \subfigure[Eliminating a $D$-bubble with two vertices.]{
 \includegraphics[width=5cm]{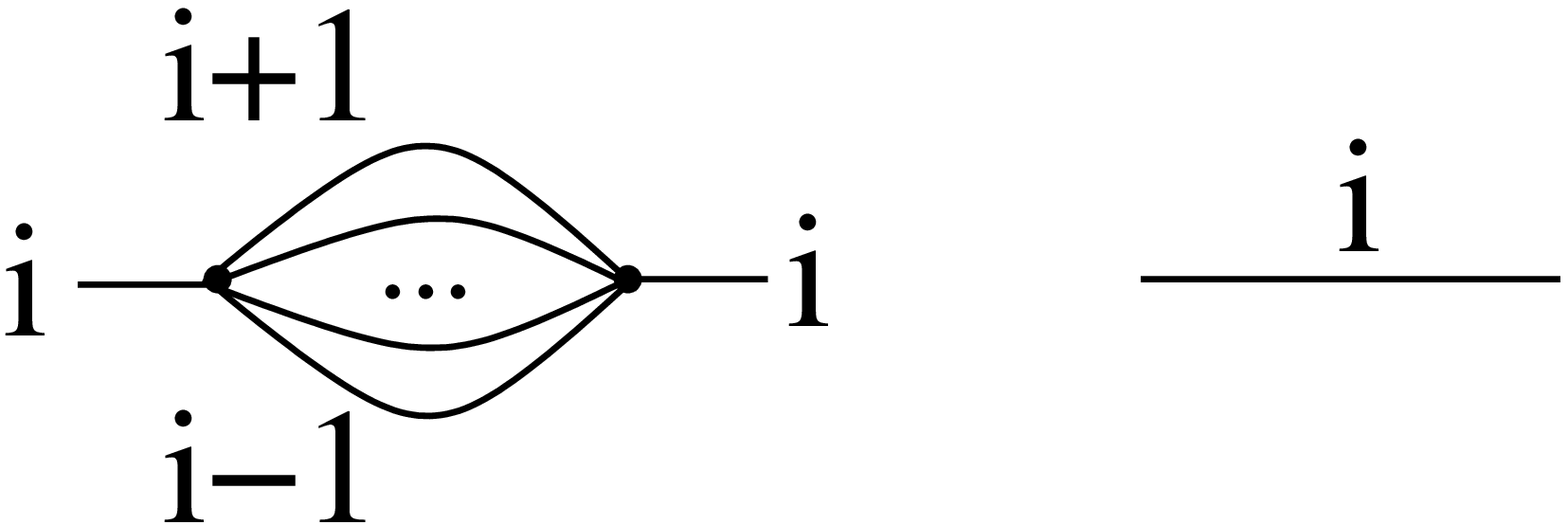}
  \label{fig:redmelon}
     } \caption{}
\end{figure}

We know from \cite{Gur4} that the leading order in the $1/N$ expansion consists in
a subclass of colored triangulations of the $D$-sphere (in \cite{Gur4}
several examples of sub-leading graphs with spherical topology are given).
Once the topology of the dominant sector is clear, we must address the combinatorics
of the dominant triangulations. Our construction relies on eliminations of $D$-bubbles
with two vertices. A $D$-bubble with two vertices
$\cB^{\widehat{i}}_{(\rho)}$ possesses $\frac{D(D-1)}{2}$ faces, hence by equation
\eqref{eq:propdegree} the degree (and the topology) of a graph $\cG$ and of the graph
$\cG_{/\cB^{\widehat{i}}_{(\rho)}}$  obtained by replacing $\cB^{\widehat{i}}_{(\rho)} $
with a line of color $i$ (see figure \ref{fig:redmelon}) are
identical\footnote{This elimination is a $1$-Dipole contraction for one of the two
lines of color $i$ touching $\cB^{\widehat{i} }_{(\rho)}$ \cite{Gur4}.
It the terminology of \cite{Gur4}, all degree zero vacuum graphs
reduce through a sequence of 1-dipole contractions to the ``super-melon'' graph depicted in the figure
\ref{fig:supermelon}.}.

We will first show that for $D\ge 3$ a leading order graph must possess a $D$-bubble with exactly two vertices.
Eliminating this bubble, we obtain a leading order graph having two less vertices. The new graph must in turn possess
a bubble with two vertices, which we eliminate, and so on. It follows that the leading order vacuum graphs must reduce
to the ``super-melon'' graph of figure \ref{fig:supermelon} after a sequence of eliminations of $D$-bubbles.
We will build a one-to-one correspondence between every leading order graph and an abstract tree
encoding this elimination process.

One can take the reversed point of view and start with the super-melon graph (for vacuum graphs)
or with the 2-point graph with a single $D$-bubble with two vertices (for the 2-point function). It is designed
to be the largest melon which will then acquire insertions of melons. At first step one chooses for each line
an integer and inserts on the line the corresponding number of $D$-bubbles with two vertices.
These bubbles form the first generation of melons. At the second step, one repeats the process by inserting $D$-bubbles
next to each other on every line, to form the second generation of melons and so
on\footnote{To our knowledge the family of melons appears for the first time in the literature
in \cite{de Calan:1981bc}.}. Of course this insertion procedure preserves colorability, degree and topology.

This leads us to define the set of {\bf melons} $\cM$ of a graph $\cG$ as the set
of 1-particle irreducible (1PI) amputated 2-point sub-graphs of $\cG$. The intuitive picture is that a
melon is itself made of melons within melons. We will see in the section \ref{sec:trees}
that the set of melons of a given graph has a natural partial ordering. The $D$-bubble with only
two vertices is obviously the smallest melon. For the 1PI amputated 2-point function the largest melon is
the graph itself.

We start by proving the first assertion in our construction. We start with lemma \ref{lemma:face-2vertices},
which probes the relationship between faces and vertices (and makes transparent why the melons
are {\bf not} the only dominant graphs in $D=2$, but are a feature of $D\geq3$). Then,
making crucial use of the jackets, we prove proposition \ref{lemma:melon}.

\begin{lemma} \label{lemma:face-2vertices}
 If $D \ge 3$ and $\cG$ is a vacuum graph with degree 0, then $\cG$ has a face with exactly two vertices.
\end{lemma}

\noindent{\bf Proof:} Since it is of degree zero, the graph $\cG$ has $ \cF = \left(\frac{D (D-1)}{2} p + D\right) $
faces, from equation \eqref{eq:propdegree}. Denote $\cF_s$ the number of faces with $2s$ vertices (every face must have
an even number of vertices). Then
\be \label{sumF_s}
\cF_1 + \cF_2 + \sum_{s\geq3}\cF_s =  \frac{D (D-1)}{2}\, p + D \; .
\ee
Let $2 p^{ij}_{(\rho)}$ be the number of vertices of the $\rho$-th face with colors $\{i,j\}$. We count
the total number of vertices by summing the numbers of vertices per face
\be \label{eq1}
\sum_{\rho, i<j} p^{ij}_{(\rho)} = \cF_1 + 2 \cF_2 + \sum_{s\ge 3} s\ \cF_{s}\;.
\ee
On the other hand, each vertex contributes to $D(D+1)/2$ faces,
$\sum_{\rho, i<j} p^{ij}_{(\rho)} \,=\, \frac{D(D+1)}{2}\, p$. We solve \eqref{sumF_s} for the number $\cF_2$ of faces
 with four vertices, and insert the result in \eqref{eq1} to get
\be
\cF_1 = 2 D + \sum_{s\ge 3} (s-2) \cF_{s} +  \frac{D(D-3)}{2}\, p \; .
\ee
Notice that on the right hand side, the first two terms yield a strictly positive contribution
for any $D\geq 2$, whereas the third term changes sign when $D=3$. Thus we
conclude that
\be
\cF_1 \geq 1\qquad \text{if $D\geq3$}\; .
\ee

\qed

Note that one can build explicit counterexamples if $D=2$.

\begin{proposition} \label{lemma:melon}
 If $D \ge 3$ and $\cG$ is a leading order 2-point graph, then it contains a $D$-bubble with exactly two vertices.
\end{proposition}

\noindent{\bf Proof:} We build the graph $\tilde \cG$ obtained from $\cG$ by reconnecting the external lines (say of color $q$)
into a new (dashed) line. It is
necessarily a leading order vacuum graph, hence it has degree $0$.
\begin{figure}[htb]
 \centering{\includegraphics[width=6cm]{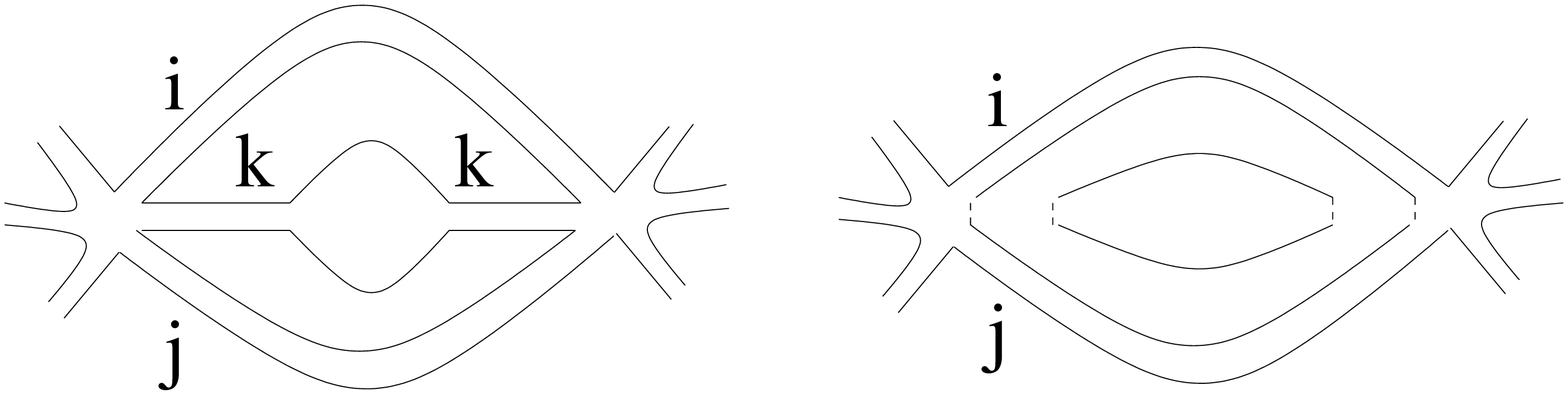}}
  \caption{The jacket $\cJ$.}
  \label{fig:fig1}
\end{figure}
It follows from the previous lemma that $\tilde \cG$ has a face with two vertices, $\cF_1$, say of colors $\{i,j\}$.
We consider the jacket $\cJ$ of $\tilde \cG$ associated to a cycle $(\dotsc, i,k,j, \dotsc)$ for some $k$. From
the proposition \ref{planar jacket}, the jacket $\cJ$ is planar, that is $\chi(\cJ)=2-2g_\cJ=2$.
We delete the two lines of color $k$ touching the lines $i$ and $j$ (see figure \ref{fig:fig1}) and
get a ribbon graph $\cJ'$. We have
\bea
 \chi(\cJ') = \chi(\cJ) +2 =4\;,
\eea
thus $\cJ'$ has two planar connected components, hence $\cG$ is two particle reducible for any couple of lines touching
$ij$. It follows that $\tilde \cG$ has the form of figure \ref{fig:fig2}, where the two point graphs $\cG^i$ and $\cG^j$ are empty.
\begin{figure}[htb]
 \centering
\subfigure[The graph $\cG$.]{
 \includegraphics[width=2.5cm]{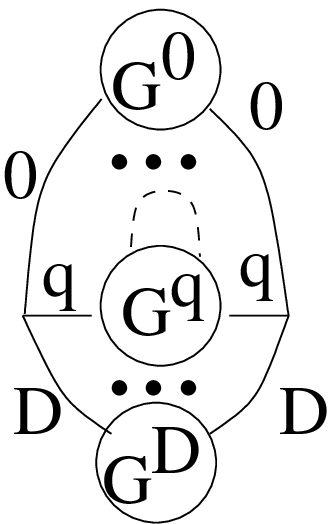}
  \label{fig:fig2}
     }
   \hskip1cm
 \subfigure[The graphs $\cG^k$ and $\tilde \cG^k$.]{
  \includegraphics[width=4cm]{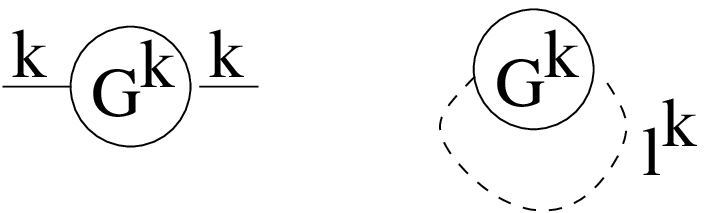}
  \label{fig:fig3}
  }
    \hskip1cm
  \subfigure[The graph $\tilde\cG^k$.]{
  \includegraphics[width=3cm]{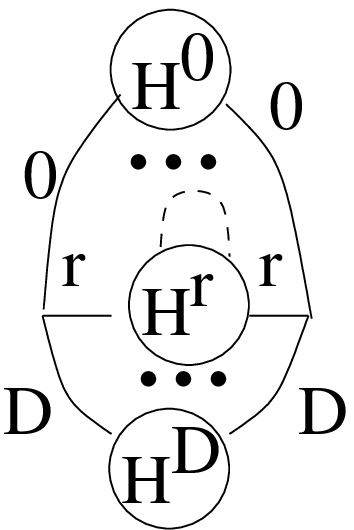}
   \label{fig:fig4}
  } \caption{}
\end{figure}
If all the two point graphs $\cG^k$, $k\neq q$ are empty then the $D$ lines of colors $k\neq q$ form a $D$ bubble with
two vertices and we conclude.

If there exists a non empty two point graph $\cG^k$, with $k\neq q$, we denote $\tilde \cG^k$ the graph obtained
by reconnecting the two external half-lines (of color $k$) into a new ``dotted'' line $l^k$ of color $k$
(see figure \ref{fig:fig3}). All sub-graphs of $\tilde \cG^k$ which do not contain $l^k$ are sub-graphs of $\cG$.
Any jacket of $\cG^k$ is a planar graph with one face broken by two external lines, hence the jackets of
$\tilde \cG^k$ (obtained from the ones of $\cG^k$ by reconnecting the two external half-lines) are planar.
It follows that $\tilde \cG^k$ is a graph of null degree $\omega(\tilde \cG^k)=0$, having at least two less
vertices than $\cG$. But $\cG^k$ contains again faces with exactly two vertices, hence takes the form in figure
\ref{fig:fig4}, where we denoted its two point sub-graphs by $\cH^t$. The line $l^k$ belongs to the two point
sub-graph $\cH^r$ (which can be trivial, i.e. formed only by the dotted line $l^k$ if $r=k$). If all $\cH^t$ for
$t\neq r$ are empty, then they form a $D$ bubble. If not, then one of them, say $\cH^s$ is not empty and has at least
two less vertices then $\cG^k$. As $\cH^s$ does not contain the line $l^k$, all its sub-graphs are sub-graphs of $\cG$.
Iterating we obtain the proposition.

\qed

\section{Counting melons via trees} \label{sec:trees}

The combinatorics of tensor models is a very difficult problem as one not only has to deal with topologies in
dimensions $D\geq3$ but also, for a given topology, one has to evaluate the number of compatible triangulations.
The $1/N$ expansion in colored tensor models classifies graphs into classes
taking into account both the topological and cellular structure of triangulations \cite{BS3}. This allows one
to access analytically the critical behavior of the leading order.

\subsection{From melons to trees}\label{sec:meltree}

We now visit more deeply that world of melons. At leading order the free energy and the connected 2-point function
write
\begin{align}
F_{\rm melons} = \sum_{p=0}^{\infty}    F_p\, (\lambda\bar{\lambda})^p \;,\qquad
G_{\rm melons} = \sum_{p=0}^{\infty}    G_p\,  (\lambda\bar{\lambda})^p \;,
\end{align}
where $F_p$ (resp. $G_p$) is the number of vacuum (resp. $2$-point) melonic graphs in dimension $D$ with $2p$
vertices. We will denote $g \equiv \lambda \bar \lambda$. The first orders in $p$ can be evaluated by a direct
counting of Wick contractions, $G_0 = G_1 =1$, $G_2=(D+1)$ etc. .

To go further we map leading order 2-point connected graphs (with external legs of color say $D$)
to a well-known species, namely {\bf $(D+1)$-ary trees} which are colored rooted trees of coordination
 $(D+2)$, \cite{manes-kary-trees}. The basic idea is
that the dominant graphs are generated by random insertions of $D$-bubbles with two vertices.

\begin{figure}[htb]
 \includegraphics[width=5cm]{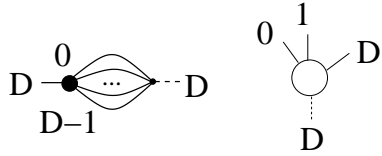}
 \caption{First order.}
  \label{fig:order1}
\end{figure}

{\bf Order $(\lambda\bar \lambda)$:} The lowest order graph consists in exactly one $D$-bubble with two vertices
(and external lines of color $D$). There is only one Wick contraction
leading to this graph, hence $G_1 =1$. We represent this graph by the tree with one vertex decorated with $(D+2)$
leaves (a leaf is a vertex of degree 1), one of them being chosen as the root. The root and a second leaf have lines of color
$D$, and the other leaves have colors $j\neq D$. On $\cG$, we consider ``active'' all lines of colors
$j\neq D$ and the line of color $D$ touching the vertex $\lambda$. They correspond to the
active leaves of the vertex (of colors $0,\dots D$). The root leaf corresponds to the external
 line of color $D$ touching $\bar \lambda$ and is inactive. See figure \ref{fig:order1},
where the vertex $\lambda$ is dotted and the inactive line and leaf are represented as dashed.

\begin{figure}[htb]
 \includegraphics[width=5cm]{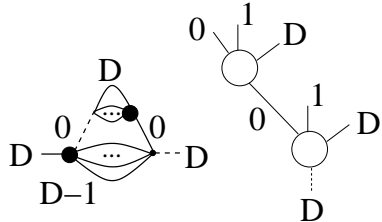}
 \caption{Second order.}
  \label{fig:order2}
\end{figure}

{\bf Order $(\lambda \bar \lambda)^2$:} At second order we have $D+1$ graphs contributing.
They come from inserting a $D$-bubble with two vertices on any of the $D+1$ active lines of the
first order graph. All the interior lines of the new $D$-bubble are active, and so is the exterior
line touching its vertex $\lambda$.
Each of those graphs has a combinatorial weight $\frac{1}{2!^2}$ and is produced by $2!^2$ Wick contractions (corresponding
to the relabelling of the vertices $\lambda$ and $\bar \lambda$), thus an overall factor $1$.
Say we insert the new bubble on the active line of color $j$. This graph corresponds to a
tree obtained from the first order tree by connecting its leaf of color $j$ to a new vertex with
 degree $(D+2)$. This new vertex has $(D+1)$ leaves, one of each color.
We count $D+1$ distinct trees, hence $G_2=D+1$ (see figure \ref{fig:order2} for the case $j=0$).

{\bf Order $(\lambda \bar\lambda)^{p+1} $:} We obtain the graphs at order $p+1$ by inserting a $D$-bubble
with two vertices on any of the active lines of a graph at order $p$. The interior lines (and the exterior line
touching the vertex $\lambda$) of the new bubble are active. We represent this by connecting a vertex of coordination
$D+2$, with $D+1$ active leaves, on one of the active leaves of a tree at order $p$. The new tree line inherits the color
of the active line on which we inserted the $D$-bubble. Each of these graphs has a combinatorial weight
$\frac{1}{(p+1)!^2}$ and is produced by $(p+1)!^2$ Wick contractions (corresponding to the relabelling of
the vertices $\lambda$ and $\bar \lambda$), thus an overall combinatorial weight $1$.

At order $(\lambda \bar\lambda)^p$ we obtain contributions (with combinatorial weight $1$) from all
rooted colored trees with $p$ vertices of degree $D+2$ and with $Dp+1$ leaves.
Such trees go under the name of $(D+1)$-ary trees in the mathematical literature \cite{manes-kary-trees, heubach-catalan}.
Basically, any vertex has either $(D+1)$ or $0$ children (with respect to the natural order starting from the root).
 Notice from the way we associate a tree to a graph, we get colored version of Gallavotti-Nicolo (GN) \cite{Gal} trees.

A rooted tree is canonically associated to a partial order. The partial ordering corresponding
to the tree we have introduced is an ordering on melons (2-point 1PI amputated sub-graphs of $\cG$) $\cM$
\bea \label{ordered melons}
 \cM_1 \geq \cM_2 \quad \text{ if } \ \begin{cases}
         \text{either}\quad \cM_1 \supset \cM_2, \\
         \text{or} \quad \left\{ \begin{aligned}
         &\exists \;  \cN_{(\rho)},
         \; \cM_1 \cup \bigl( \cup_{\rho} \cN_{(\rho)} \bigr) \cup \cM_2
          \text{ is a 2-point }\\
         & \text{  amputated connected sub-graph of $\cG$ }\\
         & \text{ with external points } \bar\lambda \in \cM_1 \text{ and }
         \lambda \in \cM_2, \end{aligned}\right.
                          \end{cases}
\eea
and $\ge$ is transitive.

\begin{figure}[htb]
\begin{center}
 \includegraphics[width=10cm]{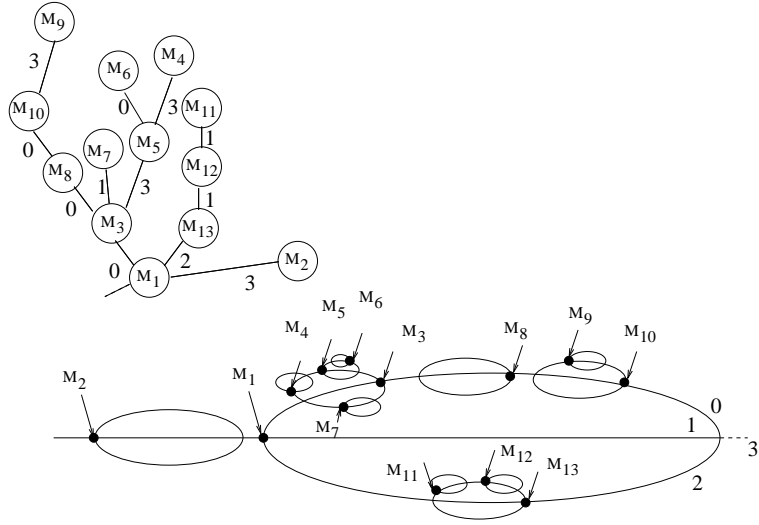}
\caption{A melon graph and its associated colored GN rooted tree.}
\label{fig:t5}
\end{center}
\end{figure}

The line connecting $\cM$ towards the root on the tree (i.e. going to a greater melon) inherits the
color of the exterior half-lines of $\cM$. An example in $D=3$ is given in
figure \ref{fig:t5} where the dotted vertices of $\cG$ are $\lambda$. The external lines
of $\cG$ are represented as dashed lines.

The first case in \eqref{ordered melons} obviously corresponds to $\cM_2$ being a sub-melon of $\cM_1$. The
second case corresponds to successive melons, say $\cM_1,\dotsc,\cM_k$, which are pairwise connected via
their external half-lines, all of the same color, say $i$. It is necessary to order them. Observe that
 $\cup_{l=1}^k \cM_l$ is a connected amputated 2-point sub-graph of $\cG$ and hence has a natural orientation due
 to the fact the two external vertices have different couplings. We order them by setting as
root of the branch formed by the melons $\cM_l$ the one containing the external point
$\bar\lambda$ of $\cup_{l=1}^k \cM_l$.

Consider the example in the figure \ref{fig:t5}, where the active leaves are implicit. We identify the melons
by their external point $\lambda$. Since the active external line of a melon is always chosen to be
the one touching the vertex $\lambda$, the root melon in an arbitrary graph is the one containing the external
point $\bar\lambda$, e.g. $\cM_1$ in figure \ref{fig:t5}. Note that $\cM_3\supset \cM_4,\cM_5,\cM_6,\cM_7$,
hence it is their ancestor. Also $\cM_3 \cup \cM_8 \cup \cM_{10}$ forms a two point function with external point
$\bar\lambda \in \cM_3 $ and as $\cM_9 \subset \cM_{10}$, the melon $\cM_3$ is the ancestor of
$ \cM_4,\cM_5,\cM_6,\cM_7, \cM_8,\cM_9,\cM_{10}$.

\subsection{Resumming the dominant series}

\subsubsection{Direct solution}

\begin{figure}[htb]
\begin{center}
 \includegraphics[width=9cm]{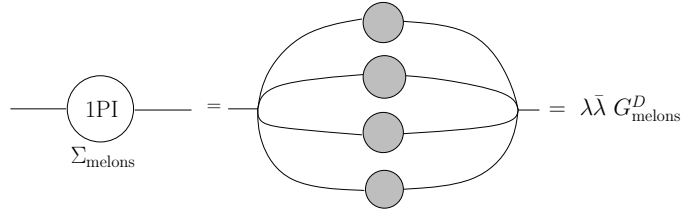}
\caption{The melonic structure is responsible for a special relation between the 1PI 2-point function and
the full 2-point function. To build the 1PI function, one just has to insert the full function on each line
of a $D$-bubble with two vertices.}
\label{fig:melonsig}
\end{center}
\end{figure}

Let $\Sigma_{\rm melons}$ be the melonic 1PI 2-point function. As it can be obtained by
arbitrary insertions of melons into melons it necessarily has the structure represented in figure
\ref{fig:melonsig}. Taking into account the equations \eqref{eq:iid} and \eqref{eq:2point}, $\Sigma_{\rm melons}$ writes as a
function of the two point connected function at melonic order $G_{\rm melons}$ as
\bea
\Sigma_{\rm melons} &=& \frac{(\lambda\bar{\lambda})}{N^{D(D-1)/2}}
\sum_{n \notin \vec n_D, p\notin \vec p_D }
\prod_{i=0}^{D-1} \langle \bar \psi^i_{\vec n_i} \psi^i_{\vec p_i} \rangle =
\frac{(\lambda\bar{\lambda})}{N^{D(D-1)/2}} \sum_{n \notin \vec n_D, p\notin \vec p_D }
\crcr
&=&\frac{(\lambda\bar{\lambda})}{N^{D(D-1)/2}}
\sum_{n \notin \vec n_D, p\notin \vec p_D }
\prod_{i=0}^{D-1} \delta_{\vec n_i \vec p_i} (G_{\rm melons})^D
 = (\lambda\bar{\lambda}) \delta_{\vec n_D \vec p_D} (G_{\rm melons})^D \; .
\eea

One gets a closed equation for $G_{\rm melons}$ by recalling that the sum of the geometric series
of 1PI amputated two point functions yields the connected two point function
$G_{\rm melons} = (1-\Sigma_{\rm melons})^{-1}$. Hence,
\be \label{melonic eq}
G_{\rm melons} = 1+ (\lambda\bar{\lambda})\,(G_{\rm melons})^{D+1}.
\ee
This equation is well-known in the literature (\cite{goulden-combinatorial} exercise 2.7.1,
\cite{graham-concrete} pp. 200, \cite{stanley-enumerative} proposition 6.2.2), specifically in various problems of enumeration
\cite{heubach-catalan}. The solution which goes to 1 when $(\lambda\bar{\lambda})$ goes to zero can
be written as a power series in $(\lambda\bar{\lambda})$ with coefficients the $(D+1)$-Catalan numbers.

\begin{proposition} \label{prop:melonic 2pt}
The melonic 2-point function admits the following expansion:
\be \label{D+1 catalan}
G_{\rm melons}(\lambda, \bar{\lambda}) = \sum_{p=0}^\infty C^{(D+1)}_p\ (\lambda\bar{\lambda})^p,
\quad \text{with}\qquad C^{(D+1)}_p = \frac1{(D+1)p+1}\,\binom{(D+1)p+1}{p}\;,
\ee
the $(D+1)$-Catalan numbers.
\end{proposition}

The usual Catalan numbers correspond to the case $D=1$. Note that the case $D=3$ can be solved explicitly.
Denote
\be
v= \frac{g^{1/3}}{ 2^{1/3} } \Big{[} \Big{(}1+\sqrt{ 1 - \frac{2^8}{3^3} g   } \Big{)}^{1/3} +
\Big{(} 1 -\sqrt{1 - \frac{2^8}{3^3} g } \Big{)}^{1/3} \Big{]},\quad \text{with}\qquad g =
(\lambda\bar{\lambda}) \;,
\ee
then
\be \label{G 3D}
G_{\rm melons|D=3} = \frac{  (1+4v)^{1/4} - \bigl[2-(1+4v)^{1/2} \bigr]^{1/2}  }{2 (vg)^{1/4}} \;.
\ee

\subsubsection{Tree-counting}

\begin{figure}
\begin{center}
 \includegraphics[width=9cm]{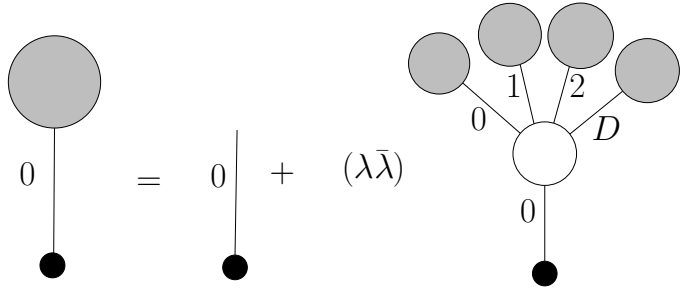}
\caption{ \label{fig:eq_trees} This picture represents the melonic equation \eqref{melonic eq} from
the point of view of the colored rooted trees. The root is depicted with a small black disk, and the
generating function is depicted as a gray disk.}
\end{center}
\end{figure}

From section \ref{sec:meltree} we conclude that $G_{\rm melons}(\lambda\bar{\lambda})$ is the
generating function of rooted $(D+1)$-ary trees. The equation \eqref{melonic eq} encodes
their proliferation.  Its interpretation in term of trees is straightforward, and depicted in the figure
\ref{fig:eq_trees}. Although the result is well known in the literature, we now give a
combinatorial proof of proposition \ref{prop:melonic 2pt}.
The two point function expands in $G_{\rm melons} = \sum_p G_p (\lambda\bar{\lambda})^p$,
with $G_p$ the number of rooted colored trees with $p$ (unlabelled) vertices of coordination $D+2$,
$Dp+1$ active leaves, and one inactive leaf on the root.

{\bf Counting by Functional Integral.} Tree counting can be mapped into functional integrals
\cite{Galla, Abdess}. Before proceeding note the following combinatorial identity
\bea\label{eq:comb}
 \sum_{k=0}^p \frac{r}{nk+r} \ \binom{nk+r}{k} \quad \frac{s}{n(p-k)+s}\ \binom{n(p-k)+s}{p-k} =
 \frac{r+s}{np+r+s} \binom{np+r+s}{p} \; ,
\eea
which is a consequence of the equations (5.58) and (5.60) in \cite{graham-concrete}.

Consider $D+1$ variables denoted $\bar \psi^0, \dotsc, \bar \psi^D$ and an unique variable $\psi$.
We define the covariance (that is a $ (D+2) \times (D+2)$ matrix)
\bea
  C(\psi,\psi)=0 \; ,\quad C(\bar \psi^i,\bar \psi^j)=0 \; ,\quad C(\bar\psi^i,\psi)  = 1 \; .
\eea
The moments of the Gaussian measure of covariance $C$ compute trivially
\bea \label{eq:mom}
 \int d\mu_C \quad  \prod_{i=0}^{D} \bigl( \bar \psi^{i} \bigr)^{n_i} \psi^m = \delta_{m, \sum n_i} m! \; .
\eea
Consider the partition function and the connected 1-point function depending of two coupling
constants, $g$ and $J$ defined as
\bea
 Z(g,J) = \int d\mu_C \quad e^{\psi J + g \psi \prod_i \bar\psi^i} \; ,
\qquad
\langle \bar\psi^D\rangle_c(g,J) =\frac{1}{Z(g,J)} \int d\mu_C \quad \bar \psi^D \; e^{\psi J + g \psi \prod_i \bar\psi^i}
\; .
\eea
The 1-point function evaluates in terms of Feynman graphs. Due to the normalization by $Z(g,J)$, all the graphs are
connected. They all possess an external half-line of color $D$, $\bar \psi^D$. A graph is made of
two categories of vertices
\begin{itemize}
 \item  the $g\, \psi \prod_i \bar\psi^i$ vertex is $(D+2)$-valent, and it has a half-line $\psi$ and $D+1$
        half-lines $\bar \psi^i$ bearing a color $i$.
 \item  the $J \psi$ vertex has only one half-line $\psi$.
\end{itemize}
Lines always connect a $\psi$ and a $\bar \psi^i$ half-lines.
Any graph of $\langle \bar\psi^D\rangle_c(g,J)$ is built in the following way.
The external half-line $\bar\psi^D$ is the root leaf (of degree 1) and must connect to a vertex. If the
latter is a $J \psi$ vertex then
the graph has exactly one line. If it is a $g \psi \prod_i \bar\psi^i$ vertex, then
each of the half-lines $\bar \psi^i$ must connect to a vertex. Each of these vertices is either a leaf
(a vertex $J\psi$, of coordination $1$), or a $(D+2)$-valent vertex $g \psi \prod_i \bar\psi^i$ with colored half-lines.
A graph can not form any loops (the end half-lines are always $\bar \psi$), and it is connected. Therefore any
graph is necessarily a tree with colored lines. Up to the relabelling of the vertices, there is exactly one Wick contraction
leading to every colored rooted tree, with $p$ vertices of coordination $D+2$ and $Dp+1$ leaves of coordination one, and one
root leaf. Hence
\bea
 \langle \bar\psi^D\rangle_c(g,J) = \sum_{p} C_p \; g^p J^{Dp+1}\Rightarrow G_{\rm melons} =
\langle \bar\psi^D\rangle_c(\lambda\bar\lambda,1) \; .
\eea

The partition function can be evaluated explicitly using eq. \eqref{eq:mom}
\bea
 Z(g,J) &=& \int d\mu_C \quad e^{\psi J + g \psi \prod_i \bar\psi^i}  =
  \sum_{p,q} \frac{g^p}{p!} \frac{J^{q}}{q!}\ \int d\mu_C \quad \bigl(\psi \prod_{i=0}^D \bar \psi^i \bigr)^p \psi^q
  \crcr
   &=& \sum_{p,q} \frac{g^p}{p!} \frac{J^{q}}{q!}\ \delta_{(D+1)p,q+p} (q+p)! =
   \sum_{p} \binom{(D+1)p}{p} \; g^p J^{Dp} \; ,
\eea
and the connected 1-point function is
\bea
&& \langle \bar\psi^D\rangle_c(g,J) =  \frac{1}{Z(g,J)} \int d\mu_C \quad \bar \psi^D \; e^{\psi J + g \psi \prod_i \bar\psi^i}
 = \frac{1}{Z(g,J)}
\sum_{p,q} \frac{g^p}{p!} \frac{J^{q}}{q!} \int d\mu_C \quad \bar \psi^D \bigl(\psi \prod_{i=0}^D \bar \psi^i \bigr)^p \psi^q
\crcr
&&= \frac{1}{Z(g,J)} \sum_{p,q} \frac{g^p}{p!} \frac{J^{q}}{q!} \delta_{(D+1)p+1,q+p} (q+p)!
 = \frac{1}{Z(g,J)} \sum_{p} \binom{(D+1)p+1}{p} \; g^p J^{Dp+1} \; .
\eea
To prove proposition \ref{prop:melonic 2pt} we must just show
\begin{remark}\label{rem:count}
 As power series,
 \bea
  \sum_{p} \frac1{(D+1)p+1}\,\binom{(D+1)p+1}{p} \; g^p J^{Dp+1} =
\frac{\sum_p \binom{(D+1)p+1}{p} \; g^p J^{Dp+1}}{ \sum_{p} \binom{(D+1)p}{p} \; g^p J^{Dp} } \; .
 \eea
\end{remark}
\noindent{\bf Proof:} Build the Cauchy product of series
\bea
 && \Big{[}  \sum_{p} \frac1{(D+1)p+1}\,\binom{(D+1)p+1}{p} \; g^p J^{Dp+1} \Big{]}  \times
\Big{[} \sum_{p} \binom{(D+1)p}{p} \; g^p J^{Dp} \Big{]} \crcr
 &&= \sum_p g^p J^{Dp+1} \; \sum_{k=0}^p \frac{1}{(D+1)k+1}\,\binom{(D+1)k+1}{k} \;
\binom{(D+1)(p-k)}{p-k} \; .
\eea
Shifting the argument of the first binomial coefficient, the sum over $k$ rewrites
\bea
 \sum_{k=0}^p \frac{1}{Dk+1}\,\binom{(D+1)k}{k} \;
\binom{(D+1)(p-k)}{p-k} \; ,
\eea
which is symmetric in $k$ and $p-k$, thus equates
\bea
&& \frac{1}{2} \sum_{k=0}^p \Big{(}\frac{1}{Dk+1}+\frac{1}{D(p-k)+1 } \Big{)}\,\binom{(D+1)k}{k} \;
\binom{(D+1)(p-k)}{p-k} \crcr
&&= \frac{Dp+2}{2} \sum_{k=0}^p \frac{1}{Dk+1} \binom{(D+1)k}{k} \; \frac{1}{D(p-k)+1 } \binom{(D+1)(p-k)}{p-k}
\crcr
&& = \frac{Dp+2}{(D+1)p+2}\binom{ (D+1)p+2}{ p } = \binom{ (D+1)p+1}{ p }\; ,
\eea
where we used eq. \eqref{eq:comb}.

\qed

{\bf Counting using Cayley's theorem.}
Denote $T_p$ the number of ordinary trees with labelled vertices with
$p$ vertices of coordination $D+2$, and $Dp+2$ vertices of coordination $1$.
They are counted by Cayley's theorem: the number ordinary trees on $n$ labelled
vertices with degree $d_i$ at vertex $i$ is $(n-2)!/\prod_i (d_i -1)! $,
hence
\be
T_p = \frac{\bigl[ (D+1)p\bigr] !}{\bigl[ (D+1)!\bigr]^p}\;.
\ee

There are two differences with respect to the trees we want: $i)$ the vertex-labelling, $ii)$ the line-coloring.
We first start with coloring the tree. The color of the inactive leaf of the root is fixed by the external color of
the 2-point function. We have $(D+1)!$ colorings of the lines of the root. Any vertex of coordination $D+2$
will support $(D+1)!$ colorings of its lines (as the color of the line connecting it towards the root is fixed by its
ancestor). Thus we pick up a factor $[(D+1)!]^p$ for the colorings of the lines. The relabellings of the vertices
bring an $\frac{1}{p!}$ for the vertices of coordination $D+2$ and an $\frac{1}{(Dp+1)!}$ for the active leaves.
We conclude that
\be
G_p = \frac{\bigl[ (D+1)!\bigr]^p}{p!\,(Dp+1)!}\ T_p = \frac{1}{p!} \frac{1}{(Dp+1)!} \; [(D+1)p]! = C_p^{(D+1)}\; .
\ee
Note that there is one subtle point in using the Cayley's theorem: one can not first relabel the vertices and then color the
lines. Indeed, in order to identify the allowed colorings of the lines one must have distinguished vertices (otherwise one can
not properly count for instance the colorings of leaves touching the same vertex).

\section{Critical behavior and continuum limit} \label{sec:critical}

\subsection{Critical behavior}

We now have the exact counting of melonic graphs in arbitrary dimensions. In particular,
we can look at the large $p$ behavior of $G_p$. From applying the Stirling's formula on the proposition \ref{D+1 catalan},
we get the following
\begin{proposition} \label{prop:entropy}
The number of melonic graphs for the $D$-sphere with $(2p)$ vertices has the following asymptotic behavior,
\be
G_p \sim A\ g_c^{-p}\ p^{-3/2}\;,
\ee
with
\be
g_c = \frac{D^D}{(D+1)^{D+1}},\quad \text{and}\qquad A = \frac{e}{\sqrt{2\pi}}\,\frac{\sqrt{D+1}}{D^{3/2}}\;.
\ee
In particular, $G_p$ is exponentially bounded by $e^{-p\ln g_c}$, with $\ln g_c<0$.
\end{proposition}

The $D$-dependent constant $g_c$ is the critical value of the coupling $g\equiv \lambda\bar{\lambda}$. Indeed,
 it is well-known that a series with coefficients going like $g_c^{-p} p^{-\alpha}$ behaves in the neighbourhood
of $g_c$ like $(g_c-g)^{\alpha -1}$, hence the most singular part of the melonic 2-point function is
\be
G_{\rm melons, sing} \sim K\ \left(\frac{g_c -g}{g_c}\right)^{1/2}\;,
\ee
for some constant $K$. This can be checked explicitly from the closed formula in the $D=3$ case, \eqref{G 3D}.

As for the free energy of melonic graphs, it is obtained thanks to the proposition \ref{relation GF}. Indeed,
the relationship between $G_N$ and $F_N$ induces the same relation at all orders in the large $N$ expansion.
 Thus, we get for the singular part of the melonic free energy
\be
G_{\rm melons} = 1 + \lambda\,\partial_{\lambda} F_{\rm melons} \quad
\Rightarrow\quad F_{\rm melons, sing} \sim K'
\ \left(\frac{g_c -g}{g_c}\right)^{2-\gamma_{\rm melons}},\text{ for}\quad \gamma_{\rm melons} = \frac12\;.
\ee
The exponent $\gamma$ is known as the susceptibility, or entropy exponent\footnote{Once an exponential
 bound is found for the proliferation of a species (melonic graphs with a fixed number of vertices) the
entropy exponent characterizes the polynomial part of the number of such objects, like $p^{-\alpha}$.}.

The critical behavior is a key ingredient to provide the random colored melonic triangulations with a
 continuum limit. But {\it a priori} the geometric interpretation of that continuum limit depends on
 the details of the model under consideration. In the following we use the natural interpretation of
the i.i.d. model as a generator of dynamical triangulations.

\subsection{Melons as branched polymers}

Now that we have extracted the main information from the family of melonic graphs, it is worth discussing
the physical implications of our results. The i.i.d. model is quite naturally interpreted\footnote{Tensor
models were indeed originally proposed in this context \cite{ambj3dqg}.} as a model of dynamical triangulations (DT)
\cite{ambjorn-book, ambjorn-houches94, ambjorn-revueDT, ambjorn-scaling4D, david-revueDT}. In the reminder of this
section we switch to notations more familiar in the DT literature.

A melonic graph is dual to a colored triangulation
of the $D$-sphere, and we denote $N_k$ the number of $k$-simplices (this is
obviously the number of $(D-k)$-bubbles). Then, the amplitude of a graph $\cG$ can be rewritten as
\be\label{eq:ampligraph}
A(\cG) = e^{\kappa_{D-2}N_{D-2}\, -\, \kappa_D N_D}\;,
\ee
where $\kappa_{D-2}$ and $\kappa_D$ are, denoting $g=\lambda\bar{\lambda}$
\be
\kappa_{D-2} = \ln N,\quad \text{and} \qquad
\kappa_D = \frac12\Bigl( \frac12\,D(D-1)\, \ln N - \ln (g)\Bigr)\;,
\ee

The argument of the exponential takes the form of the Regge action (discrete form of the
Einstein-Hilbert action for general relativity) on a triangulation with regular $D$-simplices
of length, say, $a$. Indeed, on a Regge discretization, the curvature is concentrated around the
$(D-2)$-simplices and measured by the deficit angle $\delta(\sigma_{D-2})$ ($2\pi$ minus
the sum of the dihedral angles hinged on the $(D-2)$-simplex $\sigma_{D-2}$) and
the total volume is measured by the cosmological term. The Regge action is
\be\label{eq:regge}
S_{\rm Regge} = \Lambda \sum_{\sigma_D} \vol(\sigma_D)
- \frac{1}{16\pi G}\sum_{\sigma_{D-2}} \vol(\sigma_{D-2})\,\delta(\sigma_{D-2})\;.
\ee
The volume of a regular $k$-simplex is $\vol(\sigma_k) = \frac{a^k}{k!}\sqrt{\frac{k+1}{2^k}}$,
and \eqref{eq:regge} takes the form \eqref{eq:ampligraph} on a regular triangulation if we identify
$N$ and $g$ in terms of the (bare) dimensionful parameters
$G, \Lambda$ and the  length $a$ as
\begin{align}
\ln N &= \frac{\vol(\sigma_{D-2})}{8G}\;, \nonumber \\
\ln g &=
 \frac{D}{16\pi G}\vol(\sigma_{D-2})\Bigl(\pi(D-1) - (D+1) \arccos \frac1{D}\Bigr) - 2\Lambda\,\vol(\sigma_D)
\equiv\, -2\,a^D\ \widetilde{\Lambda}\;. \label{log g}
\end{align}
Notice that in two dimensions, $\ln g = - 2\Lambda\,\vol(\sigma_D)$.

The large $N$ limit corresponds to $G\rightarrow 0$. Since $g$ is
kept finite, $\Lambda$ becomes large, positive, and scales like $\Lambda\sim 1/(a^2 G)$. The triangulations
we observe, which are dual to melonic graphs, have degree $0$, hence amplitude (recall that $g = \lambda\,\bar{\lambda}$)
\be
N^{-D}\,A(\cG) = g^{N_D/2} = e^{-a^D \widetilde \Lambda N_D}  =
e^{\frac{D}{32\pi G}\vol(\sigma_{D-2})\bigl(\pi(D-1) - (D+1) \arccos \frac1{D}\bigr) N_D}\, e^{-\Lambda\vol(\sigma_D)\,N_D}\;,
\ee
where we have explicitly split the contribution of the Einstein-Hilbert term and of the cosmological term.
The scalar curvature is the argument of the first
exponential (up to $1/G$ and irrelevant constants). It grows linearly with the number of $D$-simplices
and is positive (since $\pi(D-1) - (D+1) \arccos \frac1{D}>0$ for $D\geq3$).

A crucial point in DT is the convergence of the partition function. It is stated as the requirement
that the entropy grows linearly with the volume.
Here the entropy is the logarithm of the number of triangulations with a fixed volume, i.e. a fixed
 number of $D$-simplices. This criterion is satisfied for the melonic family, since this is just a
reformulation of the proposition \ref{prop:entropy} which proves the summability for $g< g_c$.

For the melonic family, the numbers of $D$-simplices and $(D-2)$-simplices are not independent.
 Indeed, they are related by
\be
N_{D-2} = \frac{D(D-1)}{4}N_D + D\;.
\ee
Hence, the large $N$ limit of the colored tensor model projects dynamical triangulations to
 curves parametrized by $g= \lambda\bar{\lambda}$
\be
\kappa_D - \frac{D(D-1)}{4}\,\kappa_{D-2} = - \frac12 \ln g > -\frac12 \ln g_c\;,
\ee
where $g_c$ is given in the proposition \ref{prop:entropy}. This gives the critical curve:
\be
\kappa_D^c(\kappa_{D-2}) = \frac{D(D-1)}{4}\,\kappa_{D-2} - \frac12 \ln g_c\;,
\ee
which is a non-trivial outcome of our analysis.

In the remaining of the section, we want to argue further that the family of triangulations we
observe corresponds to the phase of branched polymers (BP) which is well-known in DT (and in $D=2$
above the $c=1$ barrier) \cite{ambjorn-d>1, ambjorn-scaling4D,ambjorn-revueDT, david-revueDT, ambjorn-BP}.
Indeed, BP are known to dominate the regime of large positive curvature, where $\kappa_D>\kappa_D^c(\kappa_{D-2})$
for sufficiently large values of $\kappa_{D-2}$. Since $\kappa_{D-2} = \ln N$, this is what we expect from the
 large $N$ limit of tensor models, unless colors reduce the dominant family to a subset which has a different
statistical behavior than BP. But on the contrary the results of the previous sections are typical of BP,
and we think that the large $N$ limit of colored tensor models in general is a clean way to generate random BP.

Notice that in $D=2$, the family of dominant graphs is that of planar graphs for $c<1$, which does contain
 the BP (and the $2D$ melonic graphs). But the latter do not dominate in the limit of a large number of
 triangles (lots of vertices in the planar graphs): the critical behavior is driven by the whole
planar sector, which gives $\gamma_{\rm string} = -\frac12$. Beyond the $c=1$ barrier however, the dominant
family is reduced to BP, with a susceptibility exponent $\gamma_{\rm BP} = \frac12$.

In higher dimensions, the phase of large positive curvature has the statistical properties of BP, that is
of trees so that the susceptibility exponent is the same, $ \gamma_{\rm melons} = \gamma_{\rm BP} =\frac12$.

Another feature of the BP phase is the Hausdorff dimension $d_H=2$. Obviously, one could get
this value for melonic graphs by using the natural distance on the trees which we have described in the
previous sections. However, the vertices of the trees correspond to melons, which are somehow non-local from the
 point of view of the triangulation. Natural choices for a notion of distance are to consider either
sequences of edges of the triangulation as paths, or to use instead sequences of lines of the graph
 (i.e. $(D-1)$-simplices of the triangulation). For a given graph (and associated triangulation),
the two notions will differ in general. However, in known cases\footnote{This result has been
brought to our attention by Jan Ambj\o{}rn.}, they turn out to lead to the same
Hausdorff dimension.

The last comparison between melonic triangulations and BP is geometric. Indeed, it has been observed
in DT that in the BP phase the number of vertices and the number of $D$-simplices grow proportionally
 $N_0 \simeq N_D/D$ (see for example \cite{david-revueDT}). The natural geometric interpretation
 is the following. Start with a triangulation of the $D$-sphere and
 iterate random sequences of $(1\rightarrow (D+1))$ moves. Such moves add a vertex to a
 $D$-simplex and split it into $(D+1)$ new $D$-simplices. This way, the triangulation gets a
new vertex together with $D$ additional $D$-simplices. Hence, asymptotically, $N_0 \simeq N_D/D$.
Notice that in the crumpled polymer phase of DT, one has in contrast $N_0\sim N_D^\delta \ll N_D$.

Melonic graphs behave again like BP, since we have the exact relation:
\be
N_0 =  \frac{N_D}{2} + D\;.
\ee
The reason why the proportionality coefficient is $1/2$ in any dimension is that melons are not obtained
just through $(1\rightarrow (D+1))$ moves\footnote{This move alone does not respect the colors.}. Instead,
one adds a $D$-bubble with only two vertices to some initial melon by first performing a $(1\rightarrow (D+1))$
move, which is immediately followed by a $(D\rightarrow 2)$ move\footnote{This second move restores the colors
 spoiled by the first one.}. Such a sequence adds one vertex to the triangulation together with only two $D$-simplices.

Clearly melons provide a better balance between $N_0$ and $N_D$ and are thus more likely to maximize $N_0$
for a fixed $N_D$ than the usual set of BP. It may be surprising that they have never been considered as
 such in DT. We think that this is because the creation of a $D$-bubble with two vertices needs two
successive moves at the same place, which is unlikely to happen in Monte-Carlo algorithms.

\subsection{Large volume limit}

That interpretation of melonic graphs as (colored) branched polymers suggests a continuum limit obtained
like in matrix models by sending $a$ to zero and $g=\lambda\bar{\lambda}$ to its critical value at
 the same time. Indeed, when $g \ll g_c$, triangulations with a
large number of simplices get exponentially suppressed by $(\frac{g}{g_c})^{N_D/2}$
and triangulations with a small number of $D$-simplices dominate
the free energy $F_{\rm melons}$. However, when $g\to g_c$, the summability of $F_{\rm melons}$ is lost due to the triangulations
with a large number of $D$-simplices. This is what we intuitively expect
as a continuum limit: a triangulation with an infinite number of simplices, whose individual sizes can be sent to zero.
Notice however that this is rather a {\em large volume limit} than a continuum limit in the usual sense of field theory, since the coupling $\kappa_{D-2}$ is sent to infinity (and thus $G$ to zero) instead of tuning it to some critical value. This is exactly like the standard large $N$ limit of matrix models for pure two-dimensional gravity, and here simply due to the fact that $\kappa_{D-2}=\ln N$.

The average volume is
\begin{align}
\langle\, \Vol\, \rangle &= a^D\,\langle N_D\rangle = 2\,a^D\ \lambda\,\partial_\lambda\ \ln F_{\rm melons}\;,\\
&= 3\,g_c \frac{a^D}{g_c - g} + \text{less singular terms}\;,
\end{align}
and must be kept finite in the continuum limit. In the
 large $N$ limit, we have already sent $G\rightarrow0$, so that we are left with a single
dimensionful parameter, $\widetilde{\Lambda}$ \eqref{log g}, which should be renormalized
when reaching the critical point. The average volume writes in terms of $\widetilde{\Lambda} $
\be
\langle\, \Vol\, \rangle \sim -\frac{3\,\ln g}{2\,\widetilde{\Lambda}}\,\left(\frac{g_c - g}{g_c}\right)^{-1}\;
= \frac{3}{2\,\widetilde{\Lambda}_{\rm R}} \; , \qquad
\widetilde{\Lambda}_{\rm R} = \frac{\widetilde{\Lambda}}{-\ln g}\ \left(\frac{g_c - g}{g_c}\right)\;.
\ee
In order to obtain a continuum phase with finite physical volume we need to take the continuum limit
\be
\nonumber a\,\rightarrow\,0,\quad g\,\rightarrow\,g_c,\quad
\text{with $\widetilde{\Lambda}_{\rm R} \sim a^{-D}(g_c-g)$ fixed}\;.
\ee
Notice that this renormalization of $\widetilde{\Lambda}$ has already appeared in the DT literature,
like in \cite{ambjorn-revueDT} where it was given the equivalent form:
\be
a^D\,\widetilde{\Lambda}_{\rm R} = \kappa_D - \kappa_D^c(\kappa_{D-2})\;.
\ee

\section{Conclusion}

In this paper we have identified the structure of the dominant graphs in the large $N$ limit of the
i.i.d. colored tensor model in arbitrary dimension, as featuring couples of simplices glued along all
but one of their faces as the elementary subgraph. Such graphs can be mapped to trees, hence
they can be exactly counted, their series is summable, and a critical behavior is observed.

From the dynamical triangulation point of view, we have shown that the entropy grows linearly with
the volume and that a continuum limit is reached by tuning the coupling to its critical value, so
that triangulations with many simplices become important, while the lattice spacing is sent to
zero.  We have argued that this phase corresponds to the well-known branched polymers.

An important issue concerning the critical behavior we have observed is to understand its
universality, i.e. the relevance of the details of the microscopic model. In matrix models
for $2D$ gravity, one can describe the surfaces using either triangulations or quadrangulations.
Actually, any interaction with a positive coupling falls in that same universality class. It is
not yet clear if the colored tensor models can support similar higher degree interactions.
However, an interesting feature
revealed in our analysis is a universality with respect to the dimension $D$. Although value of
the critical coupling depends on $D$ (and goes to zero when $D$ goes to infinity), the
susceptibility exponent does not. This is in agreement with the almost systematic appearance
of a branched polymer phase in dynamical triangulations.

Although we have restricted our analysis to the i.i.d. model, the same melonic graphs dominate more
involved models like the colored Boulatov-Ooguri model \cite{GurRiv}. In fact, as melonic graphs have
the maximal number of faces at fixed number of vertices, they are likely to generically dominate
in tensor models, irrespective of the details of the covariance (propagator).

Since we have mapped melons to trees, it is appealing to think that they inherit all the
statistical properties  of trees. However, the mapping does not preserve the locality of the
triangulation, so that more investigations are needed in order to conclude whether the
physical correlations defined on the triangulation agree with those of branched polymers. In
particular, it would be interesting to test Fischer's scaling relation directly from melons.

Finally, we want to mention two possibilities to go beyond the present work in the mid-term. The first is to extend our framework to perturbations which are not around the trivial tensor, but around some non-trivial background tensor, solution of the classical equation of motion like in \cite{hamiltonian-gft}. A second option is to try to go beyond the large $N$ limit. This is necessary to reach a genuine continuum limit, where at least two couplings approach their critical values, instead of a large volume limit.

\section*{Acknowledgements}

The authors are grateful to Jan Ambj\o{}rn for his explanations on branched polymers in
matrix models and dynamical triangulations.

Research at Perimeter Institute is supported by the Government of Canada through Industry
Canada and by the Province of Ontario through the Ministry of Research and Innovation.

\end{document}